\documentclass[spanish,english]{paper}
\usepackage[T1]{fontenc}
\usepackage[latin9]{inputenc}
\usepackage{listings}
\usepackage{color}
\usepackage{babel}
\addto\shorthandsspanish{\spanishdeactivate{~<>}}

\usepackage{float}
\usepackage{amstext}
\usepackage{amssymb}
\usepackage[unicode=true,
 bookmarks=true,bookmarksnumbered=false,bookmarksopen=false,
 breaklinks=false,pdfborder={0 0 1},backref=false,colorlinks=true]
 {hyperref}
\hypersetup{pdftitle={Mapping Objects to Prolog Predicates.},
 pdfauthor={José E. Zalacain Llanes},
 pdfsubject={Mapping Objects to Prolog Predicates.},
 pdfkeywords={Prolog, Predicate, Class, Object, Relationship.}}

\makeatletter

\providecommand{\tabularnewline}{\\}

\makeatother

\begin{document}

\title{Mapping Objects to Persistent Predicates.}

\author{José E. Zalacain Llanes%
\thanks{jzalacain@nauta.cu%
}}
\maketitle
\begin{abstract}
The Logic Programming through Prolog has been widely used for supply
persistence in many systems that need store knowledge. Some implementations
of Prolog Programming Language used for supply persistence have bidirectional
interfaces with other programming languages over all with Object Oriented
Programing Languages. In present days is missing tools and frameworks
for the systems development that use logic predicate persistence in
easy and agile form. More specifically an object oriented and logic
persistence provider is need in present days that allow the object
manipulation in main memory and the persistence for this objects have
a Logic Programming predicates aspect. The present work introduce
an object-prolog declarative mappings alternative to support by an
object oriented and logic persistence provider. The proposed alternative
consists in a correspondence of the Logic Programming predicates with
an Object Oriented approach, where for each element of the Logic Programming
one Object Oriented element makes to reciprocate. The Object Oriented
representation of Logic Programming predicates offers facility of
manipulation on the elements that compose a knowledge.\end{abstract}

\section*{Introduction}

The Object Oriented Programming (OO) paradigm is see like an applications
development standard in present moment. The applications in many cases
need persistent data making of the persistence a fundamental concept.
Logic Programming (LP) with Prolog, by other hands, is a programming
paradigm with OO programming languages integration and have demonstrated
your versatility when it is used like logic persistence engine. The
data or knowledge persistence in LP take a relational character doing
than the solutions that integrate both paradigms suffer of impedance
mismatch. Considering this incompatibility, the incorporation of the
LP like persistence mechanism in object oriented solutions; take obvious
advantages permitting utilize the best of every paradigm.

Many solutions exist for communication between OO programming languages
and LP that may are utilized to give support to logic declarative
persistence in present days. SWI Prolog \cite{wielemaker:2011:tplp}
is a good alternative for the communication between the OO programming
languages and Prolog with bidirectional interfaces for languages like
C++ and Java, this last through JPL. Amzi Prolog supplies an interface
for integration with OO programming languages that can be embedded
like a logic server within application. A logic program from the Amzi
Prolog + Logic Server view is comparable with a database acceded from
procedural languages \cite{wielemaker:2011:tplp}. Another alternative
is tuProrlog \cite{denti2001tuprolog}, Prolog interpreter written
on Java for offers bi-directionality between Java and Prolog through
an inference engine that can be used like an object or several objects,
each one with a configuration and distinct knowledge base and invoking
services for the solution of questions.

Working with these solutions forces to create complex query strings
and explicitly term structures construction before querying, problem
identified by \cite{DBLP:conf/ki/OstermayerS13} for JPL case. The
queries return data structures, which must be interpreted to get the
correspondent model elements from these results. This is consequence
of the nonexistence of a strict correlation between the application
domain model and the way in that the predicates are declared. The
present paper establishes the specifications for the declarative mapping
of logic predicates to achieve a correspondence between the OO and
the PL paradigms like an alternative for application development that
require logic persistence. This initiative suggest a LP predicates
representation through the OO principal elements. The object oriented
representation of LP predicates consists in a bidirectional correlation
among the OO concepts such like class, objects, relations, and LP
terms formulated in Prolog. \\
The present paper is structured such that the first section exposes
general concepts about the PL and the OO. The section seconds exposes
the theory bases for the correspondence between the PL terms and the
OO principal elements. A study case that evidence object-declarative
mappings of Logic Predicate is offered in the third section. Finally,
the section quarter offers the conclusions.

\section*{Overview}

\subsection*{Logic Programming}

The Prolog belongs to the declarative programming languages paradigm.
Prolog is a LP language specially indicated to modeling problems that
imply objects and relations among objects \cite[p. 192]{Bratko:2000:PPA:357526}.
In Prolog, the predicates are clauses (fact or rule) which takes a
functor and arity combination \cite[p. 16]{Bramer2005}. The predicates
permit representing the relations among objects. To the most general
predicate is named relation and it is defined as the set of all predicate
instances that satisfy the relation \cite[p. 4]{Bratko:2000:PPA:357526}
like shown in equation \ref{eq:Most General Relation}. 

\begin{equation}
p(X_{1},X_{2},...,X_{n}),n\geqslant1\label{eq:Most General Relation}
\end{equation}

Variables $X_{i}$ are used to naming a term that will be determined.
When the term is determined, mean that variable is instanced or has
a substitution. Are instances of \ref{eq:Most General Relation} those
predicates for which $X_{i}$ variables find like substitutions $t_{i}$
terms. Here equation \ref{eq:Instance Relation} represents the set
of predicates particular instances of a relation like shown in \ref{eq:Most General Relation}.

\begin{equation}
p(t_{1},t_{2},...,t_{n}),n\geqslant1\label{eq:Instance Relation}
\end{equation}

Predicates are compound terms or structured data objects that begin
with an atom named functor, followed by a sequence of one or more
arguments, which are closed in parentheses and comma separated. This
arguments sequence is named objects n-tuple \cite[p. 8]{Bratko:2000:PPA:357526}.
The familiarization of this concept with another programming languages
permits representing a compound term like a structure. The functor
represents the structure name while arguments represent the fields
\cite[p. 10]{Bramer2005}.

\subsection*{Object Orientation}

The Object Orientation (OO) is seen by many peoples like a method
to organize and share code in big software systems and a technique
to organize a system in terms of objects and its relations. In the
taxonomy that offers \cite{Armstrong:2006:QOD:1113034.1113040} the
principal concepts of the OO are classified to in structural and behavioral
elements, being a part of these the concepts of Class \cite[p. 93]{Booch:2007:OAD:1407387},
Object \cite{Armstrong:2006:QOD:1113034.1113040}\cite[p. 78]{Booch:2007:OAD:1407387},
Method \cite{Armstrong:2006:QOD:1113034.1113040}, Attribute \cite[p. 290, p. 508]{Booch:2007:OAD:1407387}
and Data Type\cite[p. 65]{Booch:2007:OAD:1407387} \cite{Cook:2009:UDA:1639949.1640133}.
In OO, the classes have relations with other classes at data type
level. The most referred relations are Generalization/Specifi{}cation,
Whole/Part and Association. Generalization/Specification or inheritance
is a relation that share the structure and behavior defined in one
or more classes \cite{Armstrong:2006:QOD:1113034.1113040}\cite[p. 98]{Booch:2007:OAD:1407387}.
Whole/Part relationships have two different cases of relations between
classes. The first case is the composition where the part classes
compound the whole class and the part do not exist of independent
way. The second case is aggregation where classes are an aggregate
of the whole and they can exist of independent form \cite{Keet:2008:RRO:1412417.1412418}.
Associations relationship can be introduced in OOP languages like
attributes of participating objects in the relation or for the modeling
of the own relation like a class where the instances of the class
have the all-participating objects in the relation like his attributes
\cite{Baldoni:2007:RMR:1775223.1775255}. They are two of principal
alternatives for associations modeling, which are designated association
as attribute pattern and association as object pattern respectively
\cite{Osterbye:2007:DCL:1512762.1512769}.

\section*{Mapping Objects to Prolog Predicates}

The OO has like the principal premise the conception of than all that
he surrounds and compound the real world can be modeling like objects
or like relations among objects. Having this like principal premise
the Prolog predicates are considering of abstract way like objects.
The LP predicates representation from the OO perspective in theory
is possible because to equal than in the LP, the OO is a modeling
paradigm to describe objects and its relations. For achieve a LP predicates
representation through the OO, is fundamental see as the OO concepts
are applied at logic predicates.

\subsection*{Data Type Mapping}

Prolog is a programming language that specifies through the grammatical
syntax the different data types that manipulate. This is possible
because the language syntax specifies different forms for each data
type. All these data types are derived from an ancestor data type
named Term.  Derived Terms in Prolog are the atoms, numbers, variables
and compound terms. By other side, Prolog data types may establish
a correspondence with OO programming languages primitive data types.
OO programming languages have to Object like data type ancestor. It
is analogue to Prolog Term abstract data type ancestor. Both languages
have an especial reserved word to indicate the null element or reference
for the language. Strings types is analogue to Prolog Atom, the Floats
numbers are equivalent directly to Prolog Float data type. Integers
numbers are equivalent to Prolog Integer. The Object arrays can be
mapped directly to Prolog List of data type terms. By other side,
all Object instance of user defined classes can be mapped to Structure
compound term (Predicates) attend to some structural definitions.
In order to achieve the data type correspondence between both languages
is necessary define a mapping function where given some parameter
data type return your equivalent in the other language.

\textbf{Definition 0:} For Prolog data type set $T$ and OO language
data type set $\Theta$, is possible define a mapping function if
exist a bijective function $m:T\rightarrow\Theta$ where $\forall t\in T$,
$\exists\theta\in\Theta$, $m_{(t)}=\theta$, and have inverse bijective
function $m^{-1}:\Theta\rightarrow T$ where $\forall\theta\in\Theta$,$\exists t\in T$,
$m_{(\theta)}^{-1}=t$.

\begin{center}
\begin{table}[H]
\begin{centering}
\begin{tabular}{cc}
\hline 
\textbf{Logic Programing } & \textbf{Object Oriented }\tabularnewline
\hline 
Nil & Null\tabularnewline
True & True\tabularnewline
Fail & False\tabularnewline
Atom & String\tabularnewline
Float  & Float \tabularnewline
Integer  & Integer \tabularnewline
Structure & Object \tabularnewline
List & Array\tabularnewline
\hline 
\end{tabular}
\par\end{centering}

\centering{}\caption{Correspondence between LP data type and OO primitive data type.}
\end{table}

\par\end{center}

\subsection*{Structural and Behavioral Elements}

The conception to see a logic predicate like an abstract entity is
taken from \cite[p. 10]{Bramer2005} and being the class the principal
structural element in the OO:

\textbf{Definition 1:} For a predicate $p(X_{1},X_{2},...,X_{n}),n\geqslant1$
of which the predicates of the form $p(t_{1},t_{2},...,t_{n}),n\geqslant1$
constitute an instance, his equivalent class in the OO like his relation,
joins the common structure for all predicates with equal name and
arguments numbers. 

From implementation point view, a class correlated to a logic predicate
inherit by extension of the Object data type. Each class corresponding
to a logical predicate is a structure term in your most general form.
This conception make than any OO predicates representation defined
by the user, be an extension of the data type system in OO programing
language. The integration of all of the predicates defined by the
user to the data type system, permit than without generality loss,
these may be processed like objects. 

\textbf{Definition 2:} For a predicate $p(X_{1},X_{2},...,X_{n}),n\geqslant1$
of which the predicates of the form $p(t_{1},t_{2},...,t_{n}),n\geqslant1$
constitute an instance, his correspondent class in the OO will be
constituted with the attributes that in representation of variables
$X_{i}$ will be instanced with correspondent values of the terms
$t_{i}$. 

The attribute conception is incomplete if not talk about of the data
type. The associated data type for each attribute will be any class
that represent a data type supported by Prolog or any OO data type
that have an equivalent in Prolog data type. Independently of these
data types, an attribute may have like associated data type any class
corresponding to a predicate integrated to the data types system by
own user definition as a result of the relation modeling between classes.

\textbf{Definition 3:} For all predicates of the form $p(X_{1},X_{2},...,X_{n}),n\geqslant1$,
his correspondent class in the OO does not define a behavior for yours
objects instances in behavioral absence in logical predicates. 

The predicates declared in a logic program with this approach; do
not denote a behavior or activity visible externally. The associated
class to a logic predicate unlike another OOP class, they represent
the knowledge that define the logic predicates of abstract way. The
class corresponding to predicates only define helper methods that
allow initializing, acceding or modifying the status for each one
of the objects that will persist in the knowledge base. 

\textbf{Definition 4:} For all predicates of the form $p(t_{1},t_{2},...,t_{n}),n\geqslant1$
instance of a relation $p(X_{1},X_{2},...,X_{n}),n\geqslant1$, will
have one and only one object instance of his common class corresponding
to the most general relation that the previously mentioned predicate
belongs. 

The predicates that conform knowledge base can see like persistent
objects. The equivalent to a predicate in the OO is a particular instance
of a common class for all predicates with equal name and arguments
numbers, where the status for this object will be constituted by his
attributes values. 

\begin{center}
\begin{table}[H]
\begin{centering}
\begin{tabular}{ccc}
\hline 
\textbf{Concept} & \textbf{Logic Programming} & \textbf{Object Orientation}\tabularnewline
\hline 
Abstract entity & Relation $p(X_{1},X_{2},...,X_{n}),n\geqslant1$ & Class\tabularnewline
Property & Variable $X_{i}$ & Attribute\tabularnewline
Behavior & - & Method\tabularnewline
Entity instance & Predicate $p(t_{1},t_{2},...,t_{n}),n\geqslant1$ & Object\tabularnewline
\hline 
\end{tabular}
\par\end{centering}

\centering{}\caption{Correspondence between PL elements and OO elements.}
\end{table}

\par\end{center}

\subsection*{Relationship between Logic Programming Predicates}

Introduced the PL predicates representations through OO, where talk
about the elements such like class and objects, is essential talk
about of the relationship between predicates and his application.

\textbf{Definition 5:} For all predicates $p(X_{1},X_{2},...,X_{n}),n\geqslant1$
your equivalent class in the OO constitutes a super class for the
classes corresponding to a predicate with equal name and $k\geqslant1$
greater number of arguments. 

The specifications of a predicate $p(X_{1},X_{2},...,X_{n}),n\geqslant1$
they result in a predicate with the form shown in \ref{eq:Inheritance Relation}.

\begin{equation}
p(X_{1},X_{2},...,X_{n+k}),n,k\geqslant1\label{eq:Inheritance Relation}
\end{equation}

The present definition represent the polymorphic character of the
logic predicates. All predicate of the way shown in (3) is a predicate
$p(X_{1},X_{2},...,X_{n}),n\geqslant1$ of which inherits its structure.
The correspondent class in the model must be abstract unless this
should be instanced and declared in the knowledge base. If the super
class is abstract, the derived class will be declared like a predicate
that include the father arguments followed of the all arguments specified
by derived relation.

\textbf{Definition 6:} For all predicates $p(X_{1},X_{2},...,X_{n}),n\geqslant1$,
his correspondent class in the OO constitutes the whole in a whole/part
relation if exists a class corresponding to a predicate $q(Y_{1},Y_{2},...,Y_{m}),m\geqslant1$
such that at least $X_{i}\text{=\ensuremath{q(Y_{1},Y_{2},...,Y_{m}),m\geqslant1}}$.

Whole/Part relationship or embedded relationship between logic predicates
are identifiable when a predicate have the form shown in \ref{eq:Composition Relation}. 

\begin{equation}
p(...,q(Y_{1},Y_{2},...,Y_{m}),...),m\geqslant1\label{eq:Composition Relation}
\end{equation}

In a case like this, predicate $q$ is considered like a part of the
predicate $p$. The class correlated to $p$ will contain a reference
to the class's object $q$ but this last will not referenced to $p$.
The class correlated to the $p$ predicate knows all times the parts
that compound her, but the class correlated to the $q$ predicate
never will have a reference to the objects that he takes part. 

\textbf{Definition 7:} For $R\text{=\ensuremath{\left\{ Z_{1},Z_{2},...,Z_{m}\right\} ,m\geqslant2}}$
an arbitrary set of predicates of the form $p(X_{1},X_{2},...,X_{n}),n\geqslant1$,
it is said that his equivalent class in the OO are associated if exists
a class corresponding to a predicate \foreignlanguage{spanish}{$r(Z_{1},Z_{2},...,Z_{m})$}
such that: 

\selectlanguage{spanish}%
\begin{center}
$Z_{1}=p(X_{1},X_{2},...,X_{n_{1}})_{1}$
\par\end{center}

\begin{center}
$Z_{2}=p(X_{1},X_{2},...,X_{n_{2}})_{2}$
\par\end{center}

\begin{center}
$\vdots$
\par\end{center}

\begin{center}
$Z_{m}=p(X_{1},X_{2},...,X_{n_{m}})_{m}$
\par\end{center}

\selectlanguage{english}%
This definition proposes modeling associations between predicates
like object pattern, which consists in an object that join to all
participants in the relation. This object that join the terminal objects
of the association is referred like link, tuple or n-tuple\cite{DBLP:conf/models/DiskinD06}.
An n-tuple has a value for each association's terminal where each
value is an instance of the terminal associated class. The result
of declaring the n-tuple for association's relation the n predicates
of general form is shown in \ref{eq:Association Relation}. 

\begin{equation}
r(p(X_{1},X_{2},...,X_{n_{1}})_{1},p(X_{1},X_{2},...,X_{n_{2}})_{2},...,p(X_{1},X_{2},...,X_{n_{m}})_{m}),m\geqslant2\label{eq:Association Relation}
\end{equation}

\section*{Analysis and discussion}

Logics predicates are used for modeling relationships between objects
and for this they utilize a predicate name that identify the relationship,
followed of a number $n$ of arguments separated by comma and enclosed
in parentheses referred like objects n-tuple. By other hands the OO
association relationship are modeled using object pattern to wish
is referred like n-tuple and have a value for each attribute value
an instance of associated objects. Having these proposals like premise
is possible to deduce than every time that a logical predicate is
declared; his class in the OO represents an association between class
correspondents to the arguments of the same predicate.

The follow most general predicate declaration describe the relation
for a four side regular polygon. An identifier and an list of segments
compose the polygons in this declaration, where each segment is composed
for an identifier and the two point that define the segment. The point
relation includes an identifier and the coordinates that locate him.
The predicate include too other list of segments for refer to the
polygon diagonals. In this context 'Polygon'/3 is derived from the
most general predicate 'Poligon'/2 of which extends the identifier
and the list of segments. 'Polygon'/2 general predicate declaration
only include the identifier and the list of segments. Derived polygon
'Polygon'/3 adds missing the list of diagonal segments in the base
polygon 'Poligon'/2. The functor for structures are quoted because
they are complex atom that hold the simple class name or full class
name including namespace/package (e.g 'org.foundation.project.Polygon'). 

\begin{lstlisting}
'Polygon'(Id,Segments).
'Polygon'(Id,Segments,Diagonals).
\end{lstlisting}

From most general predicate and applying all definitions presented
in this paper, is obtained the follow Java class. The poligon base
class have the same name respect to equivalent Prolog predicate. The
'Polygon'/2 is the predicate base for the 'Polygon'/3 predicate and
like your signature suggest have two arguments the should be converted
to two class attributes. One string type atribute that hold the identifier
for the polygon and one array of Segment that hold all segments that
compound the polygon. 

\begin{lstlisting}
public class Polygon { 	
 protected String id;
 protected Segment segments[];
 public Polygon(String id,Segment[] segments) {
  this.id = id;
  this.segments = segments;
 }
}
\end{lstlisting}

In this case Tetragon class extend from Polygon append the  third
attribute which is a array of Segment that hold all segments that
constitute the diagonals. The main construcctor of Tetragon class
require id, segments and diagonals. Id and the segments will be delegated
to Polygon super class constructor while the diagonals will be setting
up in your own constructor. Tetragon class use the functor annotation
to hold the parent predicate functor. The classes names should be
the default predicate functor if no functor is specified.

\begin{lstlisting}
@functor(name="Polygon")
public class Tetragon extends Polygon {
 private Segment[] diagonals;
 public Tetragon(String id,Segment[] segments,
                           Segment[] diagonals) {
  super(id, segments);
  this.diagonals = diagonals;
 }
}
\end{lstlisting}

Create an object instance for Tetragon class require information about
your id of string type and two arrays of Segments, each Segment with
your respective Points. The classes Segment and Point are omitted.
Point class, in corresponce with 'Point'/3 predicate, have three attributes
the point id of string type and two attributes x and y of numeric
type (integers for this example). Segments class, in corresponce with
'Segment'/3 predicate, have three attributes the segment id of string
type and two attributes point0 and point1 of Point type. Objects can
be created and stored in variables and reuse them. In this example
objects are created in the act in which the tetragon object is built
to achieve better visual correspondence between elements of both languages.

\begin{lstlisting}
Tetragon tetragon = new Tetragon(
 "abcd",
 new Segment[] {
  new Segment("ab",new Point("a",2,2),new Point("b",2,6)),
  new Segment("bc",new Point("b",2,6),new Point("c",6,6)),
  new Segment("cd",new Point("c",6,6),new Point("d",6,2)),
  new Segment("da",new Point("d",6,2),new Point("a",2,2))
 },
 new Segment[] {
  new Segment("ac",new Point("a",2,2),new Point("c",6,6)),
  new Segment("bd",new Point("b",2,6),new Point("d",6,2))
 }
);
\end{lstlisting}

Throuhgt some Object-Prolog Converter the result of convert previous
object instance is a ground predicate instance of most general relation
beforely presented. Object-Prolog Converter not only convert primitive
data types. This mechanism resolve user defined data types inclusive
where the inheritance is present like the example presented. The result
structure is the same respect to most general predicate but the variable
are instantiated. 

\begin{lstlisting}
'Polygon'(
 abcd,
 [
  'Segment'(ab,'Point'(a,2,2),'Point'(b,2,6)),
  'Segment'(bc,'Point'(b,2,6),'Point'(c,6,6)),
  'Segment'(cd,'Point'(c,6,6),'Point'(d,6,2)),
  'Segment'(da,'Point'(d,6,2),'Point'(a,2,2))
 ],
 [
  'Segment'(ac,'Point'(a,2,2),'Point'(c,6,6)),
  'Segment'(bd,'Point'(b,2,6),'Point'(d,6,2))
 ]
).
\end{lstlisting}

The OO representation of logic programming predicates offers facility
of manipulation on the elements that compose a knowledge base using
some persistence provider mechanism in operations such like save,
delete and query knowledge. A logical persistence provider that implement
the subject presented in the present work would permit supporting
objects-declarative mapping of the logic-programming predicates. From
software engineering perspective, a proposal like this would make
possible the system implementation through the conceptual application
domain modeling using the Prolog language like Domain Specification
Language. A disadvantage of this approach in an OO model that describe
a big structural complexity produce predicates of the PL with big
syntactic complexity. A high level of nested objects inside another
objects produce logic predicate declarations of hard human interpretation.

\section*{Conclusions}

In present work was presented a developmental alternative for systems
that require logical persistence with a completely OO approach. The
aim of this work is establish a correlation between the OO principal
structural elements and the PL terms considering that all that composes
a knowledge base can be treated like objects. In the discussion, have
been demonstrated that the OO approach of the PL predicates constitutes
a viable alternative for the system development that requires persistence
in logic declarative form. As continuation of this work will intends
design and implement, a logical persistence provider library that
be useful for the object-declarative mapping mentioned in present
work. This logical persistence provider is considered like an application
interface provider over a Prolog inference machine for performance
the bidirectional conversion between class and logic predicates in
operations to save, to load and to query knowledge.

%\bibliographystyle{plain}
%\bibliography{motpp}

\begin{thebibliography}{99}

\bibitem{Armstrong:2006:QOD:1113034.1113040} Armstrong, D. J. The quarks of object-oriented development. Commun. ACM, ACM, 2006, 49, 123-128.

\bibitem{Baldoni:2007:RMR:1775223.1775255} Baldoni, M.; Boella, G.; Van Der Torre, L. Relationships meet their roles in object oriented programming. Proceedings of the 2007 international conference on Fundamentals of software engineering, Springer-Verlag, 2007, 440-448

\bibitem{Booch:2007:OAD:1407387} Booch, G.; Maksimchuk, R.; Engle, M.; Young, B.; Conallen, J.; Houston, K. Object-oriented analysis and design with applications, third edition. Addison-Wesley Professional, 2007

\bibitem{Bramer2005} Bramer, M. Logic Programming with Prolog Springer, 2005

\bibitem{Bratko:2000:PPA:357526} Bratko, I. Prolog programming for artificial intelligence Addison-Wesley Longman Publishing Co., Inc., 2001

\bibitem{Cook:2009:UDA:1639949.1640133} Cook, W. R. On understanding data abstraction, revisited SIGPLAN Not., ACM, 2009, 44, 557-572

\bibitem{denti2001tuprolog} Denti, E.; Omicini, A.; Ricci, A. tuProlog: A light-weight Prolog for Internet applications and infrastructures Practical Aspects of Declarative Languages, Springer, 2001, 184-198

\bibitem{DBLP:conf/models/DiskinD06} Diskin, Z.; Dingel, J. Mappings, Maps and Tables: Towards Formal Semantics for Associations in UML2 MoDELS, 2006, 230-244

\bibitem{Keet:2008:RRO:1412417.1412418} Keet, C. M.; Artale, A. Representing and reasoning over a taxonomy of part-whole relations Appl. Ontol., IOS Press, 2008, 3, 91-110

\bibitem{DBLP:conf/ki/OstermayerS13} Ostermayer, L.; Seipel, D. A Prolog Framework for Integrating Business Rules into Java Applications 9th Workshop on Knowledge Engineering and Software Engineering (KESE), 2013

\bibitem{wielemaker:2011:tplp} Wielemaker, J.; Schrijvers, T.; Triska, M.; Lager, T. SWI-Prolog Theory and Practice of Logic Programming, 2012, 12, 67-96

\bibitem{Osterbye:2007:DCL:1512762.1512769} Østerbye, K. Design of a class library for association relationships Proceedings of the 2007 Symposium on Library-Centric Software Design, ACM, 2007, 67-75


\end{thebibliography}

\end{document}